# Revisiting Atomic Patterns for Elliptic Curve Scalar Multiplication Revealing Inherent Vulnerability to Simple SCA

Alkistis Aikaterini Sigourou[1], Zoya Dyka[1,2], Sze Hei Li[1,2], Peter Langendoerfer[1,2] and Ievgen Kabin[1]
[1] *IHP – Leibniz Institute for High Performance Microelectronics,* Frankfurt (Oder), Germany
[2] *BTU Cottbus-Senftenberg,* Cottbus, Germany
{sigourou, dyka, li, langendoerfer, kabin}@ihp-microelectronics.com

*Abstract*— Elliptic Curve Scalar Multiplication denoted as $kP$ operation is the basic operation in all Elliptic Curve based cryptographic protocols. The atomicity principle and different atomic patterns for $kP$ algorithms were proposed in the past as countermeasures against simple side-channel analysis. In this work, we investigated the resistance of a $kP$ algorithm implemented in hardware using Longa's atomic patterns. We analysed its simulated power trace. We show in the example of our $kP$ implementation for the NIST EC P-256 that the field squaring operations are distinguishable from the field multiplications even if they are performed by the same field multiplier, due to the addressing of the second multiplicand. This inherent vulnerability of atomic patterns can be successfully exploited for revealing the scalar $k$.

## I. Introduction

The Elliptic Curve Cryptosystem (ECC), introduced by Victor S. Miller [1] and Neal Koblitz [2] has become a cornerstone of modern cryptographic systems due to its ability to provide high levels of security with relatively small key sizes. The fundamental operation in ECC is the multiplication of a point $P$ on the elliptic curve by a scalar $k$, called Elliptic Curve (EC) scalar multiplication and denoted as $kP$ operation. This operation is critical for the security of EC-based protocols: the analysis of the $kP$ execution(s) is the primary target for various cryptographic attacks aiming to reveal the secret scalar $k$ which we denote here further as key.

One significant class of attacks on $kP$ is side-channel analysis (SCA) attacks, which exploit different measurable physical parameters as a kind of leakages, for example power consumption or electromagnetic emissions of cryptographic implementations, to extract secret information. Among these, power analysis attacks, first demonstrated by Paul Kocher in 1998 [3], have shown particular effectiveness. Power analysis attacks can be broadly categorized into horizontal attacks, which analyze a single trace, and vertical attacks, which use multiple traces to uncover information. Simple Power Analysis (SPA) attacks against implementations of binary $kP$ algorithms belong to the horizontal attacks and exploit the distinguishability of the power consumption patterns corresponding to the processing a key bit value '1' in comparison to a key bit value '0'.

Atomicity is a well-known countermeasure principle against SPA attacks. Different Atomic patterns $kP$ algorithms were proposed in the past for example in [4], [5], [6], [7] but only few were implemented, especially in hardware [8]. So far, the resistance to SPA of hardware accelerators implementing $kP$ algorithms using atomic patterns [4], [5], [6] was not investigated experimentally.

Investigations described in [8] show that the atomic patterns proposed by [7] are vulnerable to horizontal, i.e. single-trace, address-bit SCA attacks at least if implemented in hardware. The reason of the vulnerability is the key-dependent addressing of the registers, which is an inherent part of all atomic patterns.

In this paper we implemented in hardware a binary $kP$ algorithm using the Longa's atomic patterns for the EC point doublings and additions [5] with the goal to investigate their distinguishability analysing simulated power traces.

Our analysis demonstrated that the *addressing* of the second multiplicand in the field squaring and field multiplication is a marker that can be exploited to distinguish these two operations. This holds true even if both operations are performed by the same field multiplier. This is due to the fact that the energy consumed by squaring operations is significantly less than by multiplications due to the *addressing* of the second multiplicand. Please note that this effect is only 1 clock cycle long, is observable before the clock cycles referring to the calculation of partial products, i.e. it does not depend on the implemented multiplication formula. This inherent vulnerability of atomic patterns can be successfully exploited for revealing the secret scalar $k$ attacking Longa's atomic patterns as well as other atomic patterns, too.

This paper is structured as following. Section II presents the theoretical background on atomic patterns and provides a literature review of the relevant research. Section III covers the algorithm's implementation, our kP accelerator design, and the power trace simulation of the design. In Section IV, we present our results and analysis, demonstrating the distinguishability of squarings and multiplications. Finally, Section V concludes the paper.

## II. Atomic Patterns as Countermeasure against SPA

To counteract SPA attacks, the power consumption patterns of the processing a key bit value '0' and a key bit value '1' have to be indistinguishable each from other. Binary $kP$ algorithms represent the key $k$ as a binary number and process it bit-by-bit. The processing of a key bit value '0' requires calculation of an EC point doubling ($2P$). The processing of a key bit value '1' requires calculation of two EC point operations: an EC point doubling and an EC point addition ($P+Q$). The varying forms of distinguishability between EC point doublings and additions are the main reasons for successful attacks. The indistinguishability of the power patterns of the processing of different bit values of the key can be achieved by regularization or randomization strategies.

The double-and-add-always algorithm [9] and the Montgomery ladder [10] are well-known examples of regular binary *kP* algorithms performing an EC point doubling as well as an EC point addition for each key bit value, i.e. for a '0' as well as for a '1'. This strategy causes increased execution time and energy consumption. Additionally, there are many successful attacks performed demonstrating that the regularization as a single countermeasure against SCA is not effective enough.

Several algorithmic randomization techniques aiming to achieve the indistinguishability of the power shapes of the processing of key bit values '0' and '1' have been developed in the past, including:

- *data randomization*: proposed in [9], this method involves randomizing the input data involved in computations, for example the randomization of projective coordinates of the EC point *P*.

- *randomized execution of the operation sequence*: the execution order of EC point doublings and additions can be randomized within the main loop of a *kP* double-and-add-always algorithm if a unified EC point addition formula is applied for both EC point operations, for example as proposed in [11].

- *address randomization*: Introduced in [12], [13], this technique randomizes memory addresses to obfuscate the relationship between operations and memory access patterns.

In 2002 C. Gebotys published the following idea for increasing the resistance of EC designs to SPA: the power shape of an EC point addition visually looks like two sequentially executed EC point doublings [14]. As a result, attackers are unable to distinguish between point doublings and half-addition patterns. This approach not only reduces the number of dummy operations compared to the double-and-add-always *kP* algorithms, saving both execution time and energy, but also enhances resistance to Simple Power Analysis (SPA). The approach was published as the state-of-the-art countermeasure against SPA in 2004 [15].

In 2004 an approach similar to the one of [14] was proposed as a countermeasure against SPA [4]: each process, i.e. the calculation of EC point doublings as well as additions, will be represented as multiple repetitions of the same block of instructions, called an atomic block. Consequently, power patterns of *2P* and *P+Q* operations consist of a different number of atomic block patterns that look equivalent from an SCA perspective. Later, the [5], [6], [7] proposed new – optimized – atomic pattern algorithms to improve the computation efficiency of the scalar multiplication.

The important advantage of the atomic patterns representing EC point operations is the fact that due to the (theoretically assumed) indistinguishability of the atomic blocks, it is impossible to distinguish one EC point operation from the other or even to identify the start and end of these operations. Please note that the atomicity principle, as well as the regularity principle, are based on the assumption, that addressing of different registers or blocks for loading/storing of data are side-channel equivalent operations [16]. In [8] using a hardware implementation of a *kP* algorithm using Rondepierre's atomic patterns [7] it was shown that the key-dependent addressing of different registers is a distinguishability marker for *2P* and *P+Q* operations and can be exploited when analysing a single *kP* execution trace. Due to the fact that the key-dependent addressing of the registers is an inherent part of many binary *kP* algorithms, their resistance to SPA has to be revisited, at least if the algorithms are implemented in hardware.

In this paper, we focus on the investigation of the distinguishability of the *2P* and *P + Q* atomic patterns proposed by Longa [5]. We have reviewed every publication that refers Longa up to the date of the syntax of this paper. Using Google Scholar and Semantic Scholar as search engines, we identified 79 (see [17]) and 66 (see [18]) scientific works which cite Longa's work [5]. Of these, 75 are non-duplicated, accessible, and really citing [5]. Out of the 75 distinct works, 30 focus on SCA attacks and/or SCA countermeasures, whereby 20 papers cite [5] only as a state-of-the-art countermeasure against simple SCA attacks; 7 papers claim that Longa's (or others') atomic patterns are possibly vulnerable to horizontal address-bit SCA. 12 publications describe improvements, implementations and analysis of Longa's atomic pattern algorithms, or investigate another atomic pattern claiming that their results are applicable to Longa's atomic patterns as well. For example, [16] introduced the Horizontal Collision Correlation Analysis Attack (HCCA), which exploits the correlation between power shapes of two multiplications with at least one common multiplicand compared to those with different operands. The attack requires only a single *kP* execution trace and is demonstrated using an atomic pattern algorithm on 8- and 32-bit microprocessors and ECs of different sizes [4]. The potential vulnerability of Longa's atomic patterns point doubling and point addition *kP* algorithms to HCCA was also theoretically demonstrated. HCCA exploits the similarity and distinguishability in power shapes of field multiplications, making it a horizontal data-bit SCA attack.

Only in 2 works [20], [21] Longa's atomic patterns were implemented for a binary *kP* algorithm using an open-source cryptographic library. The algorithm runs on an embedded device (TI Launchpad F28379 board). A single electromagnetic trace of a *kP* execution for the NIST EC P-256 was measured and analysed by performing a simple (automated) SCA attack. Opposite to the similar work [22] demonstrating the distinguishability of Rondepierre's atomic patterns shapes, the distinguishability of Longa's atomic patterns shapes was not demonstrated due to technical limitations caused by the equipment applied for the measurements, high noise level and time-consuming separation of the atomic patterns sub-traces and their synchronization.

To summarize, it was never practically investigated if the shapes of the EC point operations *2P* and *P+Q* implemented using Longa's atomic pattern for a hardware *kP* accelerator are distinct from one another. In this work, we implemented in VHDL a left-to-right binary *kP* algorithm for the NIST EC P-256 using the Longa's atomic patterns for EC point doublings and additions. We synthesised the *kP* design for the IHP 250 nm technology [23] and simulated the power trace of a single *kP* execution with Cadence's SimVision v.15.20 - s053. The implementation details are described in the following Section.

## III. IMPLEMENTATION DETAILS

### A. kP algorithm using Longa's atomic patterns

We implemented for the NIST EC P-256 the binary double-and-add left-to-right $kP$ algorithm using Longa's atomic patterns for EC point doubling and EC point addition operations, see Algorithm 1.

---

**Algorithm 1**: Binary *double-and-add left-to-right kP* algorithm.
Inputs: $n$-bit long scalar $k=k_{n-1} k_{n-2} \ldots k_1 k_0$; an EC point $P$

---

1. $Q_0 \leftarrow P$;
2. **for** $i = (n-2)$ **downto** 0
3.     $Q_0 \leftarrow 2 \cdot Q_0$ // Longa's atomic pattern for $2P$
4.     **if** $k_i = 1$ **then** $Q_0 \leftarrow Q_0 + P$ // Longa's atomic pattern for $P+Q$
5. Output $Q_0 = k \cdot P$

Longa's patterns follow the MNAMNAA sequence for each atomic block, where M stands for field Multiplication, N for Negation and A for Addition.

In the rest of this paper we denote EC point doubling as PD and EC point addition as PA. The PD pattern is composed of four atomic blocks $\Delta 1$, $\Delta 2$, $\Delta 3$, $\Delta 4$, and the PA pattern consists of six atomic blocks $\Delta 1, \ldots, \Delta 6$, see Table I. On the left side of each pattern in Table I the original sequence of atomic blocks and their operations is shown corresponding to [5]. On the right side of each pattern in Table I sequence of operations implemented in this work is represented. The implemented sequence of operations differs from the original due to the following reasons:

- The original sequence contains dummy operations denoted as \*; but does not mention which operation should be executed exactly, in our sequence we show all the operations we used as dummy operations.
- In our implementation we use the registers as proposed in [21] to achieve correct calculation results[1]; these registers are marked yellow in Table I.

In Table I, it is easy to see that different registers are used in the same operations of PD and PA atomic patterns. For example, in OP3 in $\Delta 1$, $T_1$ will be added to $T_4$ and the result will be stored in $T_5$ in the PD pattern, while in the PA pattern only register $T_d$ will be accessed. Similarly, in OP4 in $\Delta 1$, in the PD pattern a squaring of $T_2$ will be calculated storing the result in $T_6$, while in the PA pattern the values from two different registers – $T_x$ and $T_4$ – will be multiplied and stored in $T_5$. Other examples are OP5…OP7 in $\Delta 1$, OP1 in $\Delta 2$, etc. These examples illustrate how different registers and operations are employed in the PD and the PA patterns despite similar processing steps. All these differences can be potentially a marker to distinguish PD and PA atomic patterns, and – consequently – can be exploited in a simple SCA attack. In this work, we concentrated only on the addressing of the operands for the field multiplications and squarings.

All field squaring operations are shown on light-green background in Table I, and all field multiplications with two different multiplicands are shown on the light-red background.

---

[1] In [21] authors corrected the names of registers in few operations in Longa's atomic patterns corresponding to the formula proposed by Longa in his work.

TABLE I. IMPLEMENTED ATOMIC PATTERNS

| ATOMS | OPERATIONS | PD ATOMIC PATTERN<br>Input: $P_1 = (X_1:Y_1:Z_1)$<br>$T_1 \leftarrow X_1, T_2 \leftarrow Y_1, T_3 \leftarrow Z_1$<br>Output: $2P_1 = (X_3:Y_3:Z_3) \leftarrow (T_1, T_2, T_3)$ | | PA ATOMIC PATTERN<br>Input: $P_1 = (X_1:Y_1:Z_1)$, $P_2 = (X_2:Y_2)$<br>$T_1 \leftarrow X_1, T_2 \leftarrow Y_1, T_3 \leftarrow Z_1, T_x \leftarrow X_2, T_y \leftarrow Y_2$<br>Output: $P_1+P_2 = (X_3:Y_3:Z_3:X_1':Y_1')$<br>$\leftarrow (T_1, T_2, T_3, T_5, T_5)$ | |
|---|---|---|---|---|---|
| | | Original operations [5] | Implemented operations | Original operations [5] | Implemented operations |
| $\Delta 1$ | OP1 | $T_4 \leftarrow T_3 \cdot T_3$ | $T_4 \leftarrow T_3 \cdot T_3$ | $T_4 \leftarrow T_3 \cdot T_3$ | $T_4 \leftarrow T_3 \cdot T_3$ |
| | OP2 | \* | $T_d \leftarrow -T_d$ | \* | $T_d \leftarrow -T_d$ |
| | OP3 | $T_5 \leftarrow T_1 + T_4$ | $T_5 \leftarrow T_1 + T_4$ | \* | $T_d \leftarrow T_d + T_d$ |
| | OP4 | $T_6 \leftarrow T_2 \cdot T_2$ | $T_6 \leftarrow T_2 \cdot T_2$ | $T_5 \leftarrow T_x \cdot T_4$ | $T_5 \leftarrow T_x \cdot T_4$ |
| | OP5 | $T_4 \leftarrow -T_4$ | $T_4 \leftarrow -T_4$ | $T_6 \leftarrow -T_1$ | $T_6 \leftarrow -T_1$ |
| | OP6 | $T_2 \leftarrow T_2 + T_2$ | $T_2 \leftarrow T_2 + T_2$ | $T_5 \leftarrow T_5 + T_6$ | $T_5 \leftarrow T_5 + T_6$ |
| | OP7 | $T_4 \leftarrow T_1 + T_4$ | $T_4 \leftarrow T_1 + T_4$ | \* | $T_d \leftarrow T_d + T_d$ |
| $\Delta 2$ | OP1 | $T_5 \leftarrow T_4 \cdot T_5$ | $T_5 \leftarrow T_4 \cdot T_5$ | $T_6 \leftarrow T_5 \cdot T_5$ | $T_6 \leftarrow T_5 \cdot T_5$ |
| | OP2 | \* | $T_d \leftarrow -T_d$ | \* | $T_d \leftarrow -T_d$ |
| | OP3 | $T_4 \leftarrow T_5 + T_5$ | $T_4 \leftarrow T_5 + T_5$ | \* | $T_d \leftarrow T_d + T_d$ |
| | OP4 | $T_3 \leftarrow T_2 \cdot T_3$ | $T_3 \leftarrow T_2 \cdot T_3$ | $T_7 \leftarrow T_1 \cdot T_6$ | $T_7 \leftarrow T_1 \cdot T_6$ |
| | OP5 | \* | $T_d \leftarrow -T_d$ | \* | $T_d \leftarrow -T_d$ |
| | OP6 | $T_4 \leftarrow T_4 + T_5$ | $T_4 \leftarrow T_4 + T_5$ | $T_8 \leftarrow T_1 + T_1$ | $T_8 \leftarrow T_7 + T_7$ |
| | OP7 | $T_2 \leftarrow T_6 + T_6$ | $T_2 \leftarrow T_6 + T_6$ | \* | $T_d \leftarrow T_d + T_d$ |
| $\Delta 3$ | OP1 | $T_5 \leftarrow T_4 \cdot T_4$ | $T_5 \leftarrow T_4 \cdot T_4$ | $T_9 \leftarrow T_5 \cdot T_6$ | $T_9 \leftarrow T_5 \cdot T_6$ |
| | OP2 | \* | $T_d \leftarrow -T_d$ | \* | $T_d \leftarrow -T_d$ |
| | OP3 | $T_6 \leftarrow T_2 + T_2$ | $T_6 \leftarrow T_2 + T_2$ | $T_8 \leftarrow T_8 + T_9$ | $T_8 \leftarrow T_8 + T_9$ |
| | OP4 | $T_6 \leftarrow T_1 \cdot T_6$ | $T_6 \leftarrow T_1 \cdot T_6$ | $T_4 \leftarrow T_3 \cdot T_4$ | $T_4 \leftarrow T_3 \cdot T_4$ |
| | OP5 | $T_1 \leftarrow -T_6$ | $T_1 \leftarrow -T_6$ | \* | $T_d \leftarrow -T_d$ |
| | OP6 | $T_1 \leftarrow T_1 + T_1$ | $T_1 \leftarrow T_1 + T_1$ | \* | $T_d \leftarrow T_d + T_d$ |
| | OP7 | $T_1 \leftarrow T_1 + T_5$ | $T_1 \leftarrow T_1 + T_5$ | \* | $T_d \leftarrow T_d + T_d$ |
| $\Delta 4$ | OP1 | $T_2 \leftarrow T_2 \cdot T_2$ | $T_2 \leftarrow T_2 \cdot T_2$ | $T_4 \leftarrow T_y \cdot T_4$ | $T_4 \leftarrow T_y \cdot T_4$ |
| | OP2 | $T_5 \leftarrow -T_1$ | $T_5 \leftarrow -T_1$ | $T_{10} \leftarrow -T_2$ | $T_{10} \leftarrow -T_2$ |
| | OP3 | $T_5 \leftarrow T_5 + T_6$ | $T_5 \leftarrow T_5 + T_6$ | $T_4 \leftarrow T_4 + T_{10}$ | $T_4 \leftarrow T_4 + T_{10}$ |
| | OP4 | $T_5 \leftarrow T_4 \cdot T_5$ | $T_5 \leftarrow T_4 \cdot T_5$ | $T_{10} \leftarrow T_4 \cdot T_4$ | $T_{10} \leftarrow T_4 \cdot T_4$ |
| | OP5 | $T_2 \leftarrow -T_2$ | $T_2 \leftarrow -T_2$ | $T_8 \leftarrow -T_8$ | $T_8 \leftarrow -T_8$ |
| | OP6 | $T_2 \leftarrow T_2 + T_2$ | $T_2 \leftarrow T_2 + T_2$ | $T_1 \leftarrow T_6 + T_8$ | $T_1 \leftarrow T_{10} + T_8$ |
| | OP7 | $T_2 \leftarrow T_2 + T_4$ | $T_2 \leftarrow T_2 + T_5$ | \* | $T_d \leftarrow T_d + T_d$ |
| $\Delta 5$ | OP1 | | | $T_8 \leftarrow T_2 \cdot T_9$ | $T_8 \leftarrow T_2 \cdot T_9$ |
| | OP2 | | | $T_6 \leftarrow -T_1$ | $T_6 \leftarrow -T_1$ |
| | OP3 | | | $T_6 \leftarrow T_6 + T_7$ | $T_6 \leftarrow T_6 + T_7$ |
| | OP4 | | | $T_6 \leftarrow T_6 \cdot T_{10}$ | $T_6 \leftarrow T_6 \cdot T_4$ |
| | OP5 | | | $T_9 \leftarrow -T_8$ | $T_9 \leftarrow -T_8$ |
| | OP6 | | | $T_2 \leftarrow T_6 + T_9$ | $T_2 \leftarrow T_6 + T_9$ |
| | OP7 | | | \* | $T_d \leftarrow T_d + T_d$ |
| $\Delta 6$ | OP1 | | | $T_3 \leftarrow T_3 \cdot T_5$ | $T_3 \leftarrow T_3 \cdot T_5$ |
| | OP2 | | | $T_4 \leftarrow -T_7$ | $T_4 \leftarrow -T_7$ |
| | OP3 | | | $T_4 \leftarrow T_1 + T_4$ | $T_4 \leftarrow T_1 + T_4$ |
| | OP4 | | | $T_5 \leftarrow T_4 \cdot T_4$ | $T_5 \leftarrow T_4 \cdot T_4$ |
| | OP5 | | | $T_6 \leftarrow -T_8$ | $T_6 \leftarrow -T_8$ |
| | OP6 | | | $T_6 \leftarrow T_2 + T_6$ | $T_6 \leftarrow T_2 + T_6$ |
| | OP7 | | | \* | $T_d \leftarrow T_d + T_d$ |

Fig.1 shows the sequences of squarings, i.e. the multiplications of two identical operands (denoted as S) and multiplications of two different multiplicands (denoted as M) which are coloured as it is done in Table I and demonstrate clearly that PD atomic patterns can be easily distinguished from PA atomic patterns if:

- the parts of the power trace corresponding to multiplications or squarings can be determined in the power trace, and
- squaring operations are distinguishable from multiplications.

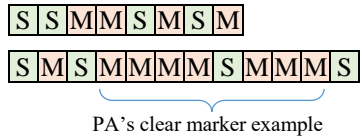

Fig. 1. Sequence of Squarings and Multiplications on PD's (top) and PA's (bottom) atomic patterns.

In section IV-*A* we demonstrate that the distinguishability of S from M is true for the in hardware implemented $kP$ designs. We use Longa's patterns as an example to discuss that the addressing of the second multiplicand if identical to the first one is the reason causing the SCA vulnerability.

### B. Our k*P* design

Our *kP* design consists of functional blocks for addition, subtraction and multiplication in the prime finite field *GF(p)* for the NIST EC P-256 [24], as well as of 256-bit long registers. The block Controller oversees the order in which field activities are carried out as well as the data loading in and out of blocks. A multiplexer enables the connection between the functional blocks and the registers. In each single clock cycle only one of the blocks or registers can load its output data to the bus. This source block and the block(s) that receive the data from the bus are determined by the Controller.

The field multiplier calculates the product of two elements of *GF(p)* corresponding to the 4-segment Karatsuba multiplication formula [25], which was adapted for multiplication of prime field elements [26]. The multiplier needs 12 clock cycles for the product calculation, starting from obtaining its first operand and finishing with the output of the field product as follows:

- $1^{st}$ clock cycle: the block Controller addresses the register/block containing the multiplicand's value for loading it to the bus; the multiplier will be addressed for reading its $1^{st}$ multiplicand from the bus;
- $2^{nd}$ clock cycle: the block Controller addresses the register/block containing the second's multiplicand's value for loading it to the bus; the multiplier will be addressed for reading its $2^{nd}$ multiplicand from the bus; at this clock cycle the value of its $1^{st}$ multiplicand will be stored to a multiplier's internal register
- $3^{rd}…11^{th}$ clock cycles: in each of these clock cycles a single partial product will be calculated, the partial products will be accumulated in a multiplier's register corresponding to the implemented multiplication formula, and the reduced result will be stored in the multiplier's output register.
- $12^{th}$ clock cycle: the multiplier finished the calculation of the product; the field product is available in the multiplier's output register; the multiplier can be addressed for the loading the calculated product value to the bus.

For the field additions and negation which is a subtraction from 0, we used the same block i.e. the Adder, which can calculate a sum or a difference of two field elements. For the calculations, the block Adder requires 3 clock cycles as follows:

- $1^{st}$ clock cycle: the block Controller addresses the register/block containing the operand's value for loading it to the bus; the block Adder will be addressed for reading its $1^{st}$ operand from the bus; this operand will be stored in an internal register of the block;
- $2^{nd}$ clock cycle: the block Controller addresses the register/block containing the $2^{nd}$ operand's value for loading it to the bus; the block Adder will be addressed for reading its $2^{nd}$ operand from the bus; at this clock cycle the value of its $1^{st}$ operand is available in the internal register of the Adder; additionally, at this clock cycle a signal determining the required operation – addition or subtraction – is active. In case of addition, the value available in the internal Adder's register (i.e. the $1^{st}$ operand value) will be added to the $2^{nd}$ operand; the result will be reduced and stored in the internal register instead of the $1^{st}$ operand. In case of subtraction, the difference of both operands will be reduced and stored in the Adder's internal register;
- $3^{rd}$ clock cycle: the block Adder finished the calculation and the result is available in its register; it can be addressed for the loading the calculated value to the bus.

In Fig. 2, we represented the MNAMNAA sequence of the operation managed by the Controller, clock by clock. Each cell corresponds to a single clock cycle, showing 12 cycles for one field Multiplication (M), and 3 cycles for each Negation (N) and Addition (A) of the field elements. For the Multiplication, the first two clock cycles (light green) represent the register addressing of the two multiplicands, followed by 9 clock cycles calculating the partial multiplication (red), and the final cycle is used to address the multiplier for the loading of its output to the bus. The 3 clock cycles long activity of the block Adder is also represented corresponding to the description given above. Negations are marked with blue rectangles and additions are marked with green rectangles. Thus, the duration of an atomic block in our implementation is 39 clock cycles.

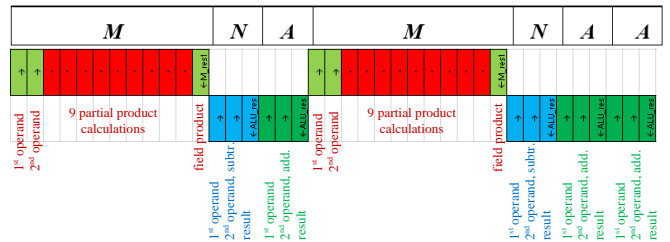

Fig. 2. Implemented sequence of operations within a single atomic block: the sequence shows the activity of functional blocks clock-by-clock; the sequence is managed by the block Controller.

In our implementation, we were aware that some calculations as well as data transmission between the design's blocks can be partially parallelized. In this work, we implemented all operations sequentially, as proposed by Longa to avoid any issue with the fact that the implementation does not correspond to Longa's original work [5]. Using Cadence's SimVision v. 15.20-s053, we synthesised the design for the IHP 250 nm cell library SGB25V [23] for a clock cycle period of 30 ns.

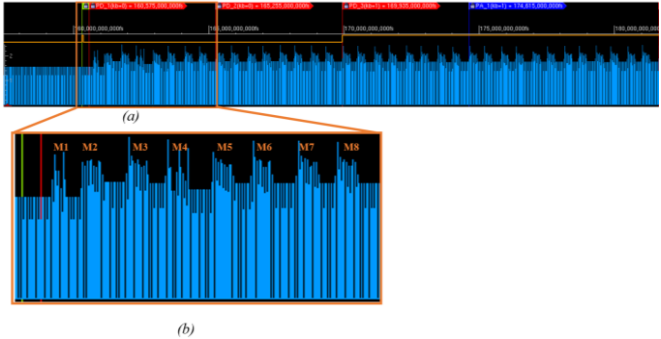

Fig. 3. *A part of the simulated power trace: (a) – the key bit values $k_{255}k_{254}k_{253}k_{252}$=1001 are processed; (b) – $k_{254}$=0 is processed performing the first EC point doubling; its power shape exposes the distinguishability of $1^{st}$ and $4^{th}$ field multiplications (see M1 and M4) from all other multiplications. The distinguishability is caused by the special operand with value 1 processed in M1 as well as in M4.*

### C. Power Trace Simulation

Using Synopsys PrimeTime v. Q-2019.12-SP1, we simulated a power trace (PT) of a *kP* execution using a time resolution of 0.1 ns, resulting into 300 samples per clock cycle. For the power trace simulation we selected a 256-bit long scalar *k* and an EC point $\boldsymbol{P}= (x, y)^2$ on the EC P-256. The PT of the simulated *kP* execution consists of 255 point doubling (PD) patterns and 134 point addition (PA) patterns. Fig. 3 shows a part of the simulated power trace visualized using SimVision.

The orange line represents the key, i.e. the scalar *k*. The blue line is the simulated PT of the *kP* execution. The most significant bit of the scalar *k* is processed before the main loop of the algorithm i.e. during the initialization phase. The corresponding time is shown with the green marker indicating the start of the *kP* operation. Corresponding to Algorithm 1, for every key bit value '0' a PD operation has to be executed, and for each key bit value '1' a PD operation followed by a PA operation has to be executed, with exception the most significant bit that acts as a secret bit and only used for initiating the *kP* execution (green marker). Using red (PD) and blue (PA) markers, we marked the start of the first operations based on the key used. In Fig. 3(a), we zoomed in on the first four atomic patterns: 3 PD patterns and 1 PA pattern.

The field multiplication is the most time and energy consuming operation in our design. Due to this fact, it is straightforward to identify the parts of the PT corresponding to all field multiplications. Each atomic block consists of 2 multiplications. Each PD pattern consists of 4 atomic blocks, i.e. 8 multiplications. Each PA pattern consists of 6 atomic blocks, i.e. 12 multiplications. In the first PD pattern, i.e. in the processing of the second most significant bit of the scalar *k*, two multiplications have to be performed with the special multiplicand value of 1 (see Table I, OP1 in Δ1 and OP4 in Δ4 of PD: after the initialization phase register T3 carries the value of Z=1). Multiplying of two operands, one of which is equal to 1, requires less energy than multiplying big operands. This fact was observed in the past and exploited for revealing the $2^{nd}$ most significant bit attacking a *kP* accelerator implementing the Montgomery ladder for the NIST EC B-233 [27]. A similar leakage was observed for ECs over prime finite fields in [28]. In our case it can be clearly seen that the $1^{st}$ and the $4^{th}$ multiplication, denoted as M1 and M4 in Fig. 3*(b)*, exhibited fewer high peaks compared to the other multiplications, where both operands were near or equal to 256 bits in length.

In this section, we demonstrated that the power shapes of all field multiplications can be clearly determined in the PT. In the next section we demonstrate that multiplications with two identical operands (i.e. squaring operations) can be distinguished from the multiplications with two different multiplicands focusing on the analysis of the power consumption in the clock cycles, in which the multiplier will be addressed for obtaining its $2^{nd}$ operand.

## IV. DISTINGUISHABILITY DEMONSTRATION AND EXPLOITING

### A. Multiplications vs. squarings: visible distinguishability

We selected the first field multiplication (OP1) in the second atomic block Δ2 to demonstrate the distinguishability of the shapes corresponding to a field multiplication with two different operands (M) compared the one of a squaring (S). In the PD pattern it is the operation $T_5 \leftarrow T_4 \cdot T_5$, i.e. it is a multiplication M. In the PA pattern it is the operation $T_6 \leftarrow T_5 \cdot T_5$, i.e. it is a squaring operation S (see Table I)[3]. We cut 12 clock cycles long sub-traces of these operations in each PD and PA in the whole *kP* trace, and overlaid them with the help of Matlab R2023a. Fig. 4 shows all sub-traces: M-sub-traces are red, S-sub-traces are blue. Additionally, the clock signal is shown (see green line above the power sub-traces), and the processes performed clock-by-clock during a multiplication are represented.

Fig.4 demonstrates clearly that power consumption in the clock cycle corresponding to obtaining the $2^{nd}$ multiplicand significantly differs for M and S. This difference can be exploited for a SPA after preparing a trace as described in the next sub-section. The reasons for the observed distinguishability can be explained as follows:

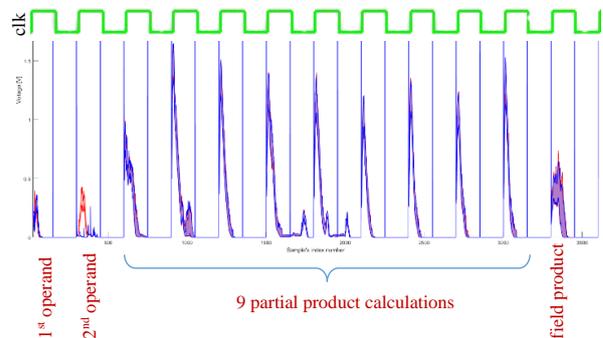

Fig. 4. *Sub-traces of the $1^{st}$ field Multiplication in the atomic block Δ2 of each EC point doubling pattern (M-sub-traces, red lines) and EC point addition patterns (S-sub-traces, blue lines).*

---

[2] The values of the scalar *k* and coordinates of the EC point *P* used in simulation, in hexadecimal:
k= 4fe342e2fe1a7f9b8ee7eb4a7c0f9e162bce33576b315ececbb6406837bf51f5

x=6b17d1f2e12c4247f8bce6e563a440f277037d812deb33a0f4a13945d898c296
y=4fe342e2fe1a7f9b8ee7eb4a7c0f9e162bce33576b315ececbb6406837bf51f5
[3] This operation corresponds to the $3^{rd}$ field multiplication in each atomic pattern, see for example M3 in Fig.3.-(b).

During the 2nd clock cycle of a field Multiplication with different operands, the Controller addresses a new register (compared to the previous cycle) to load its value onto the bus. This causes many logic cells in the muxer to switch their states, to connect the output of the newly addressed register to the input of the field multiplier. In this cycle, the muxer processes:

- A new register address compared to the previous clock cycle.
- A new value to be provided to the multiplier.

However, in the case of a squaring (S), where the second operand is identical to the first, the Controller addresses the same register, meaning neither the address nor the value changes from the previous cycle. As a result, all the logic cells of the muxer remain in the same state, and the muxer consumes no energy. Generally, muxer is a big block and makes a significant contribution to the overall *kP* design's energy consumption. This lack of energy consumption during the 2nd clock cycle of Squaring, compared to Multiplication, creates a noticeable difference in the power trace.

Thus, in the power trace of a *kP* execution, Multiplications and Squarings can be not only selected but also distinguished from each other by comparing the shape of the clock cycle corresponding to obtaining the 2nd multiplicand. If this distinction is key-dependent, these processes become – generally – a strong SCA leakage source.

### B. SPA Attack exploiting distinguishability of M and S

As we mentioned in the previous section, the *kP* power trace has to be prepared for the simple power analysis. Due to the fact that only clock cycles corresponding to the obtaining of the 2nd multiplicand will be analysed, we truncated all these clock cycles and concatenated them forming a new, power trace. A part of the new trace is shown in Fig. 5. Each field multiplication is represented there by only this clock cycle. No other operations are represented. Fig. 5*(b)* and Fig. 5*(c)* demonstrate clearly that the distinguishability of Multiplications and Squarings can be exploited to distinct the atomic patterns of the EC PD from the atomic patterns of the EC PA: the sequence of the M and S operations shown in Fig. 1 can be easily "seen". The separation of the trace in doubling and addition parts leads to the successful revealing of the processed scalar *k*.

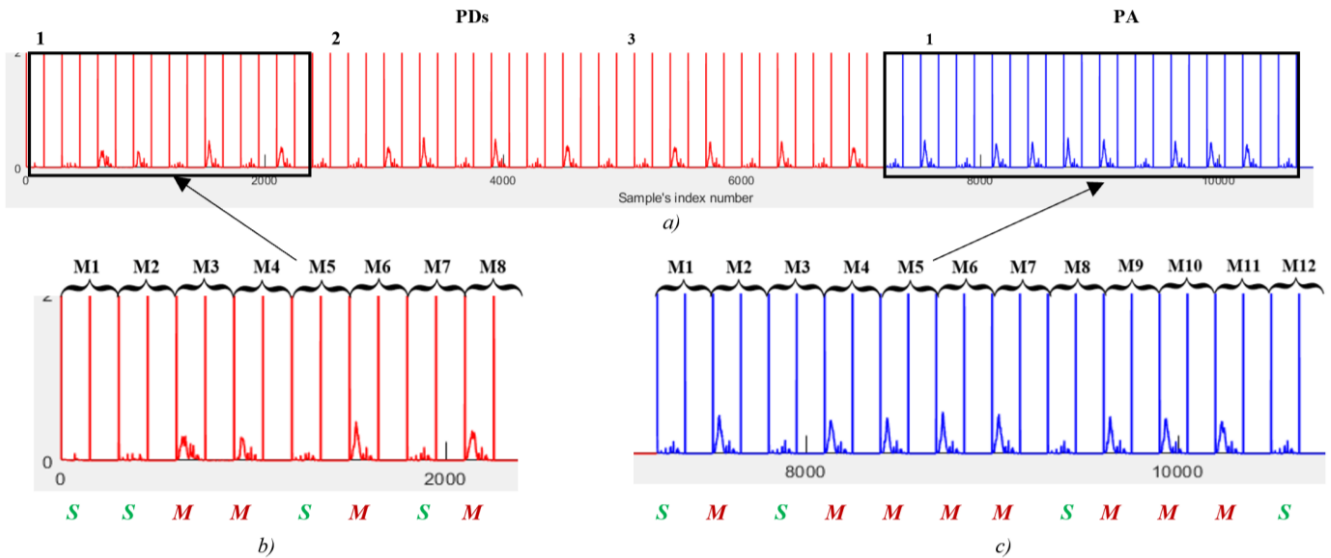

Fig. 5. A part of PT prepared for the analysis: only clock cycles corresponding to obtaining of 2nd multiplicand are represented here as a continuous sequence: *a)* – first 36 field multiplications are represented with only its 1 clock cycle corresponding to obtaining of 2nd multiplicand; *b)* – 8 field multiplications of the first EC point doubling are shown, zoomed in; *c)* – 12 field multiplications of the first EC point addition are shown, zoomed in.

In our investigations, we focused on the analysis of the field multiplications, due to the fact that no field multiplication is a dummy operation, i.e. the demonstrated vulnerability is an inherent feature of Longa's atomic block patterns and not a result of a "weak" implementation caused by a designer. Please note that the demonstrated distinguishability of M and S operations can be exploited by attacking each binary *kP* algorithm if the sequence of M and S operations is key-dependent, for example also for the MNAA atomic block patterns proposed in [4].

### V. CONCLUSION

In this research, we made two contributions: first, we described a working hardware implementation of the binary left-to-right *kP* algorithm by using the IHP 250nm technology and Longa's atomic patterns. Secondly, we pinpointed and revealed an important weakness in Longa's approach. This observation is crucial because it reveals a novel source of side-channel leakage, which can be exploited for successful SPA attacks. This holds true for each binary *kP* algorithm, in which, despite its regularity or atomicity the sequence of multiplications and squarings for an EC point doubling differs from the one for an EC point addition. The vulnerability demonstrated here has to be considered when designing ECC in hardware. Dummy operations increasing the energy consumption of *kP* designs during the transfer of the 2nd multiplicand in case of the squarings can for example reduce this SCA leakage. To the best of our knowledge, no other research has specifically identified this type of leakage based on the distinction between addressing identical and different multiplicands in elliptic curve cryptography or similar cryptographic implementations. This insight provides a new avenue for understanding and potentially mitigating side-channel vulnerabilities in cryptographic systems. Our findings highlight the need for further security enhancements in hardware cryptographic implementations to mitigate such risks.